\documentclass[preprint]{aastex}
\usepackage{lineno}
\usepackage{natbib}
\usepackage{psfig}
\usepackage{graphicx}
\usepackage{amssymb}
\usepackage{color}

\begin{document}


\shorttitle{Modeling the pulse profile of PSR J0737--3039A}
\shortauthors{PERERA ET AL.}

\title{Realistic Modeling of the Pulse Profile of PSR J0737--3039A}
\author{B.~B.~P.~Perera\altaffilmark{1,2}, C.~Kim\altaffilmark{1,3}, M.~A.~McLaughlin\altaffilmark{1}, R.~D.~Ferdman\altaffilmark{4}, M.~Kramer\altaffilmark{5}, I.~H.~Stairs\altaffilmark{6}, P.~C.~C.~Freire\altaffilmark{5}, and A.~Possenti\altaffilmark{7}}

\altaffiltext{1}{Department of Physics, West Virginia University, Morgantown, WV 26506, USA}
\altaffiltext{2}{Jodrell Bank Centre for Astrophysics, School of Physics and Astronomy, The University of Manchester, Manchester M13 9PL, UK}
\altaffiltext{3}{Department of Physics and Astronomy, Seoul National University, Seoul, 151-742, Korea}
\altaffiltext{4}{Department of Physics, McGill University, Ernest Rutherford Physics Building, 3600 University Street, Montreal, QC H3A 2T8, Canada}
\altaffiltext{5}{Max-Planck-Institut f$\ddot{u}$r Radioastronomie, Auf dem H$\ddot{u}$gel 69, D-53121 Bonn, Germany}
\altaffiltext{6}{Department of Physics and Astronomy, University of British Columbia, 6224 Agricultural Road, Vancouver,British Columbia V6T 1Z1, Canada}
\altaffiltext{7}{INAF-Osservatorio Astronomica di Cagliari, Loc. Poggio dei Pini, Strada 54, 09012 Capoterra, Italy}

\begin{abstract}
The Double Pulsar, PSR J$0737$--$3039$A/B, is a unique system in which both neutron stars have been detected as radio pulsars. As shown in Ferdman et al., there is no evidence for pulse profile evolution of the A pulsar, and the geometry of the pulsar was fit well with a  double-pole circular radio beam model.  Assuming a more realistic polar cap model with a vacuum retarded dipole magnetosphere configuration including special relativistic effects, we create synthesized pulse profiles for A given the best-fit geometry from the simple circular beam model. By fitting synthesized pulse profiles to those observed from pulsar A, we constrain the geometry of the radio beam, namely the half-opening angle and the emission altitude, to be $\sim$$30\degr$ and $\sim$10 neutron star radii, respectively.
Combining the observational constraints of PSR J0737--3039A/B, we are able to construct the full three-dimensional orbital geometry of the Double Pulsar. 
The relative angle between the spin axes of the two pulsars ($\Delta_{\rm S}$) is estimated to be $\sim$$(138\degr\pm5\degr)$ at the current epoch and will likely remain constant until tidal interactions become important in $\sim$85~Myr, at merger. 
\end{abstract}

\maketitle


\section{Introduction}

PSR J$0737$--$3039$ is the only neutron star - neutron star (NS-NS) binary in which both NSs have been detected as radio pulsars \citep{bdp+03,lbk+04}. This unique system provides an opportunity to determine the beam geometry of the individual pulsars, allowing us to construct the full three dimensional orbital and spin geometry of the binary. This information is vital in order to understand binary formation/evolution involving supernova natal kicks \citep{plp+04,fkl+11} and to study gravitational wave (GW) signals and outcomes from NS-NS mergers \citep[e.g.,][]{plp+13,ahl+08,it00}.
This system also provides the most precise test to date of general relativity in the strong-field regime \citep{ksm+06}.

The first-born, recycled pulsar of the system, PSR J0737--3039A (hereafter pulsar A), has a spin period ($P_{\rm s}$) of 22.7~ms. The second-born and slower PSR J0737--3039B (hereafter pulsar B) rotates every 2.8 s.
The two pulsars orbit each other in a tight ($P_{\rm orb}=2.4$~hrs) and moderately eccentric ($e \sim$$0.088$) orbit \citep{ksm+06}. Relativistic spin precession is expected from such binary systems, and the geodetic spin precession rates of the two pulsars are theoretically predicted to be $4\fdg8$~yr$^{-1}$ and $5\fdg1$~yr$^{-1}$ for pulsars A and B, respectively \citep[e.g.,][]{bo75b}. Relativistic precession has been measured for two binary systems based on observation: for pulsar B from a detailed study of pulsar A eclipses \citep{bkk+08} and for the NS-NS binary PSR B1534$+$12 from the observed secular and periodic variations in pulse profiles due to spin precession and aberration \citep{sta04}. A long-term pulse profile analysis of pulsar B reveals that the relativistic spin precession results in a dramatic evolution in the pulse profile, finally culminating in the radio emission disappearance with respect to our line-of-sight \citep{pmk+10}. By fitting an elliptical emission beam model, the geometry of pulsar B, including the emission altitudes, was constrained \citep{plg+12}.

Although the spin precession rates of the A and B pulsars are comparable, there is no evidence for secular variation in the pulse profile of A \citep{mkp+05,fsk+08,fsk+13}. The most plausible explanation for this stable pulse profile is that pulsar A's spin misalignment, $\delta_{\rm A}$, from the orbital normal is very small.
Assuming the two pulse components in A's profile are formed from the two radio beams corresponding to each magnetic pole, \citet{fsk+08}  determined its geometry with a double-pole circular beam model. Using six years of data, \citet{fsk+13} found that pulse profiles show no evidence for shape evolution. By fitting their model to measured pulse profile widths at different intensity levels (30\%, 35\%, 40\%, 45\%, and 50\% with respect to the peak) individually, they constrained the magnetic misalignment and the spin colatitude of the pulsar to be $\alpha_{\rm A} = 90\degr\pm 16\degr$ and $\delta_{\rm A} < 2\fdg3$ at a 68\% confidence level, respectively. This geometry supports the orthogonal configuration of pulsar A.  
 

Recent {\it Fermi} observations of the Double Pulsar revealed pulsed gamma-ray emission from pulsar A \citep{gkj+13}. The gamma-ray emission is explained by outer magnetosphere models: the ``outer gap'' (OG) model \citep{chr86a,rom96a} in which the emission is generated within the gap region between the null-charge surface and the light cylinder, or the ``two-pole caustic'' (TPC) model \citep{dr03} in which the emission is generated within the gap region extending from the NS surface to the light cylinder. \citet{gkj+13} found that the peaks of the gamma-ray and the radio profiles are not aligned in pulsar A's spin phase. This implies that the radio and gamma-ray emission are generated in different locations in pulsar A's magnetosphere. Therefore, we strongly believe that pulsar A's radio emission originates within the inner magnetosphere. They also constrained pulsar A's geometric angles $\alpha_{\rm A}$ and $\zeta_{\rm E}$ by fitting OG and TPC models separately to observed gamma-ray light curves combining a single-altitude hollow cone radio beam model to observed radio pulse profiles. In contrast, we use only radio observation in this study and model the radio pulse profile by using a radio beam model in which the emission is active across its entire beam. 
Both outer magnetosphere models derived geometry given in \citet{gkj+13} also favor an orthogonal configuration for pulsar A.

The main purpose of this work is to develop a realistic model of the radio beam geometry of pulsar A. We first repeat the analysis of \citet{fsk+13}, but analyze pulse widths including lower intensity levels such as 5\% in order to incorporate any subtle changes in pulsar A's pulse profiles.
Then, for a more realistic model, we consider a polar cap (PC) beam model that involves the dipole magnetic field structure to fully describe the shape of the pulsar magnetosphere. 
We assume that pulsar radio emission is generated at a lower altitude within the PC region, where the open field lines are located. In general, pulsar magnetosphere models are constructed at the following two limits: (a) a vacuum limit \citep{deu55} and (b) a force-free magnetohydrodynamics (MHD) limit with a plasma-filled magnetosphere \citep{spi06}.
However, a true magnetosphere operates between these two limits \citep[see][]{lst+12,kkh+12}.
The MHD solutions are considered to be more realistic, but are computationally expensive to implement \citep[see][]{hsd+08}. Further, \citet{hdm+11} found the rotating dipole magnetosphere in vacuum provides better fits to observed high-energy pulse profiles than the force-free magnetosphere. Therefore, we incorporate a semi-analytic, widely used, vacuum retarded dipole magnetic field structure \citep[e.g.,][]{yad97,dh04,bs10} with a PC radio emission model to construct the beam geometry of pulsar A.
This allows us to estimate the half-opening angle and the emission height for pulsar A.

In Section~\ref{data}, we present the data and the profile width variation of pulsar A. In Section~\ref{circular}, we constrain the pulsar geometry using the same simple circular beam model given in \citet{fsk+13} with more pulse width measurements. In Section~\ref{advanced}, we describe our analysis with the PC model based on a vacuum retarded dipole configuration. We then compare pulsar A's beam geometry obtained from the circular and PC beam models. Combining our results with those from \citet{plg+12}, we determine the orbital geometry of the Double Pulsar, including the relative angle of the spin axes of two pulsars in Section~\ref{full_geo}. Finally, we discuss results in Section~\ref{dis}.

\section{Pulse Profiles of Pulsar A}
\label{data}

In this section, we describe our analysis of pulsar A's pulse profiles. We use the same data set that \citet{fsk+13} used in their analysis (from 2005 June (MJD 53524) to 2011 June (MJD 55721) at an observing frequency of 820~MHz). All pulse profiles are constructed with 2048 bins across the spin phase, resulting in a time resolution of $\sim$$10$~$\mu$s. As shown in Figure~1 therein, pulsar A's pulse profile has not significantly changed within that time span.

\begin{figure*}
\epsscale{0.6}
\plotone{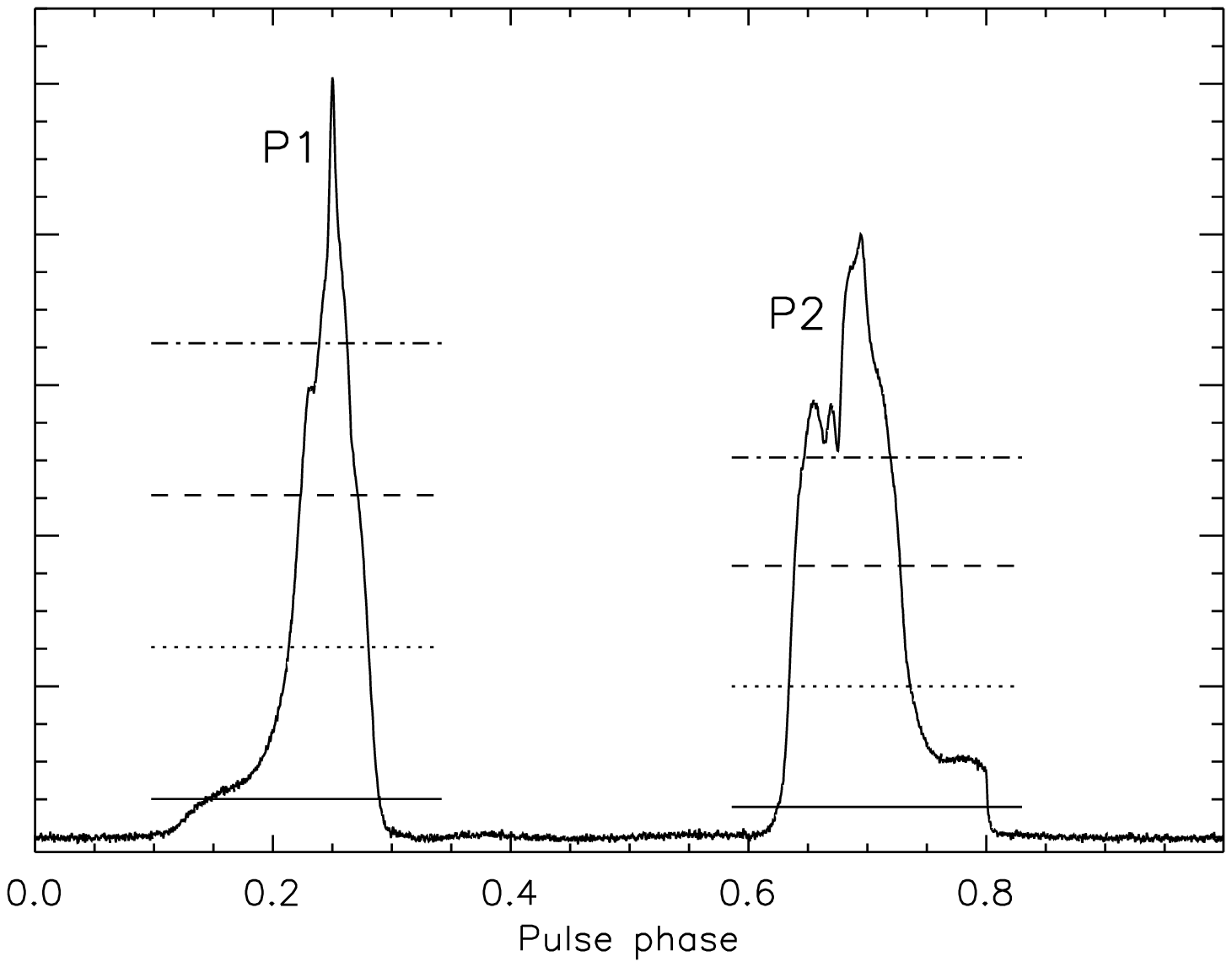}
\caption{
Integrated pulse profile of pulsar A on 6 May 2006 (MJD 53861) at 820~MHz observing frequency. The primary narrower pulse component is denoted as P1 and the secondary broader component is denoted as P2. There are 2048 bins across the spin phase. Note that the fluctuations around pulse phase 0.38 are artifacts and not real. The different intensity levels with respect to each component's peak height are marked with horizontal lines, namely 5\% ({\it solid}), 25\% ({\it dotted}), 45\% ({\it dashed}), and 65\% ({\it dotted-dashed}). The pulse widths are obtained with respect to these intensity levels. 
\label{profile}}
\end{figure*}

\begin{figure*}
\epsscale{1.0}
\plotone{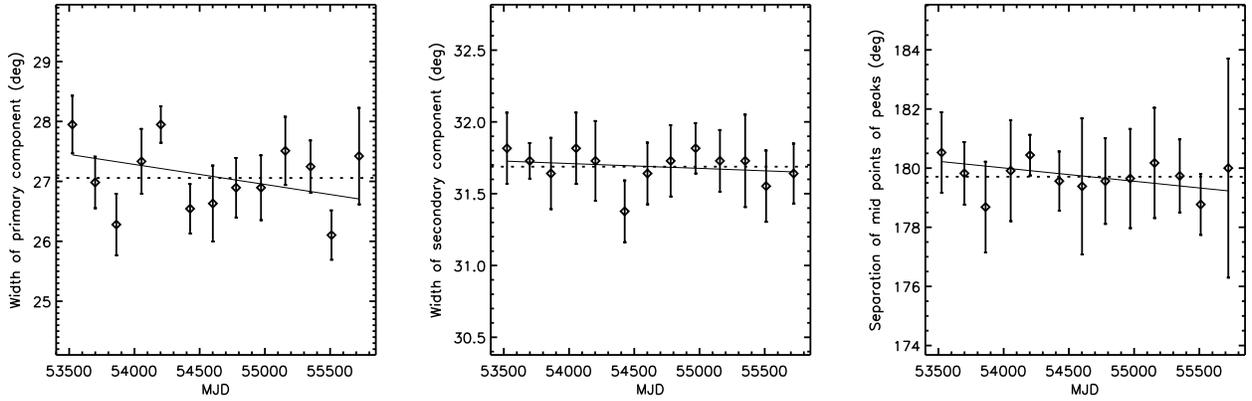}
\caption{
Pulse profile widths at 5\% of their peak heights. The errors of widths are computed from the off-pulse rms of pulse profiles. {\it Left}: Profile widths of the brightest component (P1) over time. {\it Middle}: Profile widths of the second brightest component (P2) over time. {\it Right}: The separation of the centers of the two components at their 5\% intensity level. The least-square fitting to the data is shown as a solid line. The mean value is shown as a dotted horizontal line in each panel. 
The best-fit slopes are $-(0.1\pm0.1)\degr$~yr$^{-1}$ ({\it Left}), $-(0.01\pm0.03)\degr$~yr$^{-1}$ ({\it Middle}), and $-(0.2\pm0.2)\degr$~yr$^{-1}$ ({\it Right}). 
We note that P1, P2, and the separation do not show a variation over time within their errors. 
\label{width}}
\end{figure*}

The brightest component (P1) of A's pulse profile is narrower than the secondary component (P2) (see Figure~\ref{profile}).
We note that some studies defined P2 as the brightest component \citep[e.g.,][]{fsk+13,gkj+13}. 
At each epoch, we calculate pulse widths of P1 and P2, separately, at different intensity levels (5\%, 25\%, 45\%, and 65\%) relative to each component's peak height.
The uncertainties of these widths are calculated from the off-pulse root-mean-square deviation.
We selected these particular intensity levels in order to reflect the width evolution. For instance, the 5\% width  includes any subtle changes of the profile at lower intensity, and
the other intensity levels were chosen with 20\% increments between 5\% and 65\% to avoid noise properties such as the plateau region around 10\% and the feature around 70\% of component P2 (see Figure~\ref{profile}). 
Figure~\ref{width} shows the pulse widths obtained from P1 (left panel) and P2 (middle panel) at the 5\% intensity level. 
The root-mean-square deviations of the measured widths for P1 and P2 are $0\fdg5$ and $0\fdg1$, respectively, which are smaller than the typical uncertainties of these measurements.
In addition, the least-square fits show that the 5\% widths of P1 and P2 decrease  with a rate of $0\fdg1(1)$~yr$^{-1}$ and $0\fdg01(3)$~yr$^{-1}$, respectively. Together, these indicate that there is no significant variation in pulse widths over time. This is consistent with the results  obtained  in previous studies \citep{mkp+05,fsk+13}.   
Note also
 that P1 and P2 are separated by almost $180(4)\degr$ at the 5\% level (see Figure~\ref{width}), supporting the assumption that A is an orthogonal rotator and the two pulse components are due to seeing a radio beam from each pole of the NS.

\section{Beam Geometry of Pulsar A with a Simple Circular Beam}
\label{circular}

In this section, we repeat the analysis of \citet{fsk+13} using the same simple circular radio beam model and measured pulse widths from 
P1 and P2 at different intensity levels (5\%, 25\%, 45\%, and 65\%) to constrain $\alpha_{\rm A}$ and $\delta_{\rm A}$ independently of the line-of-sight.
The only difference between our analysis and that of \citet{fsk+13} is that they calculated pairs of ($\alpha_{\rm A}$, $\delta_{\rm A}$) for each intensity level, while we obtain a single pair of ($\alpha_{\rm A}$ and $\delta_{\rm A}$) considering all four intensity levels. This results in values with smaller error bars. 
Once we obtain $\alpha_{\rm A}$ and $\delta_{\rm A}$, we fix both angles in order to estimate the half-opening angle of the beam $\rho_{\rm A}$ at the 5\% intensity level. We assume the 5\% intensity level is roughly the boundary of the beam.

Detecting a stable pulse profile over time implies that our line-of-sight always observes nearly the same cross section of pulsar A's radio beam.
Therefore, as long as one is only concerned with global geometric angles such as $\alpha_{\rm A}$ and $\delta_{\rm A}$, the circular radio beam is a simple, yet valid, choice.
Thus, we do not investigate other complex shapes such as an elliptical beam which was used for pulsar B where our line-of-sight cuts through significantly different parts of the beam in a short-timescale of years \citep{plg+12}. 

Following the analysis described in \citet{fsk+13}, 
we calculate the model-estimated pulse profile width $w_{\rm j}(t)$ at a given epoch $t$ for a given $\alpha_{\rm A}$, $\delta_{\rm A}$, $\rho_{\rm A,j}$, and $T_0$ and then fit the observed pulse width at the same epoch $t$ to the model-estimated width. 
Here, {\it j} represents different intensity levels and $T_0$ is a reference epoch when the spin axis is in the plane of the line-of-sight and the orbital normal axis.
We follow this method for measured pulse profiles at all epochs. 
By using a likelihood analysis combined with a grid search  as described in \citet{pmk+10} and \citet{plg+12}, we obtain $\alpha_{\rm A}$ and $\delta_{\rm A}$.
During the fitting procedure, we use a single $\delta_{\rm A}$ and $T_0$ for both pulse components P1 and P2, assuming A is an orthogonal rotator.

We assume that the two radio beams of the pulsar are circular and have independent beam sizes: this model is denoted as CBM. 
Therefore, we vary the north ($\rho_{\rm A,N}$) and south ($\rho_{\rm A,S}$) beams for each parameter combination ($\alpha_{\rm A}$, $\delta_{\rm A}$) until we get the maximum likelihood. 
After searching the entire parameter space, we obtain the best-fit values as follows: 
 $\alpha_{\rm A} = 88\fdg1^{+3\fdg0}_{-0\fdg6}$, $\delta_{\rm A} \leq 2\fdg8$ with a best-fit of $0\fdg9$, and $T_0 = 61800$ (see Table~1). 
The beam sizes $\rho_{\rm A,N}$ and $\rho_{\rm A,S}$ at the 5\% intensity are $27\degr \pm 1\degr$ and $32\degr \pm 1\degr$ for P1 and P2, respectively.
Note that, our best-fit $\alpha_{\rm A}$ and $\delta_{\rm A}$ are consistent with the results reported in \citet{fsk+13} within their 68\% uncertainties.

\begin{table*}
\begin{center}
\caption{Geometric parameters of known pulsar binaries with geodetic precession. $\Omega_{\rm prec}$ is the expected relativistic spin precession rate, $\alpha$ is the magnetic misalignment angle, $\delta$ is the colatitude of the pulsar's spin axis, $\zeta_{\rm E}$ is the viewing angle of the line-of-sight, $T_0$ is a reference epoch, $\rho$ is the half-opening angle of the pulsar's radio beam, and $h$ is the radio emission altitude at the edge of the beam in units of neutron star radius $R_{\rm NS}$, assumed to be 10~km. 
The 68\% error is given within parentheses; if the two limits of the error are asymmetric, the largest value is quoted.
Note that the first half of the table shows pulsar A's geometric parameters constrained based on different methods. 
The two values of $h$ correspond to those estimated for the beams from north and south magnetic poles, respectively. } 
\begin{tabular}{lcccccccc}
\hline
\multicolumn{1}{c}{PSR name} &
\multicolumn{1}{c}{$\Omega_{\rm prec}$} &
\multicolumn{1}{c}{$\alpha$} &
\multicolumn{1}{c}{$\delta$} &
\multicolumn{1}{c}{$\zeta_{\rm E}$} &
\multicolumn{1}{c}{$T_0$} &
\multicolumn{1}{c}{$\rho$} &
\multicolumn{1}{c}{$h$} &
\multicolumn{1}{c}{References} \\
 & ($\degr$~yr$^{-1}$)         & (\degr)        & (\degr)  & (\degr)   & (MJD)         &  (\degr)  & (R$_{\rm NS}$)  & \\
\hline
Pulsar A: from \\
CBM 	& 4.8             & 88(3)               & 1(2)     	& [87.8, 89.6]$^a$	& 61800   & 27(1), 32(1)    & 10(2), 11(2) & this work  \\
TPC$^b$		&    -             & 80(9)               & 0$^c$  & 86(14)		& --      & 32(1), 32(1)    & 11(2), 11(2) & this work, 1  \\
OG$^b$		&    -             & 88(17)              & 0$^c$  & 74(14)		& --      & 33(2), 38(1)    & 12(2), 15(2) & this work, 1  \\
RVM$^b$		&    -             & 99(8)               & 0$^c$  & 96(13)		& --      & 30(1), 33(1)    & 10(2), 12(3)  & this work, 1  \\
\hline
Pulsar A	& -	          & 90(8)	        & $<$6.1  	& --				& --      & $<90$     & -- &  2  \\
		&	-	  & 90(16)		& $<$2.3	& --				& --	  & 12$^d$, 19$^d$	      & -- &  3  \\
Pulsar B	& 5.1             & 61(8)               & 138(5)        & [50, 133]			& 57399   & 14.3      & [15, 38]$^e$ & 4 \\
B1913$+$16	& 1.2		  & 153(8)	        & 22(8)          & [130, 154]		& 98296   & 9      & -- &  5 \\
B1534$+$12	& 0.5            & 103(1)     		& 25(4)		& [52, 102] 		& -- 	  & 4.9      & -- &  6,7 \\
J1141$-$6545    & 1.4            & 160(16)    	        & 93(16)	& [20, 166]			& 53000   & --        & -- &  8  \\
J1906$+$0746   & 2.2	          & 81(66)		& 89(85)        &  --		& --	& --	& --    & 9 \\		
\hline
\end{tabular}
\tablecomments{
$^a$The angle $\zeta_{\rm E}$ is calculated using Equation~(4) given in \citet{fsk+13} with the best-fit parameters. \\
$^b$ The pulsar geometric angles $\alpha$, $\delta$, and $\zeta_{\rm E}$ for TPC, OG, and RVM models are taken from \citet{gkj+13}. 
The radio beam geometry $\rho$ and $h$ are estimated from our PC model (see text). \\ 
$^c$TPC, OG, and RVM results are based on $\delta_{\rm A} = 0\degr$ \citep{gkj+13}. \\
$^d$This $\rho$ is estimated for 30\% intensity level. \\ 
$^e$The limits of the emission height are given. Due to large spin misalignment of pulsar B, the emission heights corresponding to the two bright phases vary with time. \\
References: (1) \citet{gkj+13}; (2) \citet{fsk+08}; (3) \citet{fsk+13}; (4) \citet{plg+12}; (5) \citet{kra98}; (6) \citet{sta04}; (7) \citet{tds05}; (8) \citet{mks+10}; (9) \citet{dkc+13}}
\end{center}
\end{table*}

\section{Beam Geometry of Pulsar A with a Retarded Vacuum Dipole PC Model}
\label{advanced}

The circular radio beam model discussed in the previous section provides the information about the geometry ($\alpha_{\rm A}$, $\delta_{\rm A}$) of the pulsar with beam size $\rho_{\rm A}$. In order to estimate the radio emission altitudes in detail, we need to account for the magnetic field line structure. In this section, we investigate the radio emission beam of A by applying a dipole magnetosphere configuration and assume that the 5\% intensity levels of the profile, or wings, are generated from the radio emission near the boundary of the last open and closed field lines. 

Following what was derived in \citet{deu55} and used in \citet{yad97}, \citet{dh04}, and \citet{bs10}, we model pulsar A's magnetosphere by a vacuum dipole field at retarded time $t_{\rm r} = t - r/c$. For a pulsar rotating around the z-axis with an angular velocity of $\Omega$ and magnetic inclination $\alpha_{\rm A}$, the time dependent magnetic moment is given as $\vec{\mu}(t) = \mu(\sin\alpha_{\rm A}\cos\Omega t \hat{x} + \sin\alpha_{\rm A}\sin\Omega t \hat{y} + \cos\alpha_{\rm A} \hat{z})$ in Cartesian coordinates. Then the magnetic field of the retarded dipole can be written as

\begin{equation}        
\vec{B}_{\rm ret} = -\left[ \frac{\vec{\mu}(t)}{r^3} + \frac{\dot{\vec{\mu}}(t)}{cr^2} + \frac{\ddot{\vec{\mu}}(t)}{c^2r} \right] + \vec{r}\cdot \left[ 3\frac{\vec{\mu}(t)}{r^3} + 3\frac{\dot{\vec{\mu}}(t)}{cr^2} + \frac{\ddot{\vec{\mu}}(t)}{c^2r} \right]\vec{r}~,
\end{equation}

\noindent
where $r=|\vec{r}|$ is the radial distance and $c$ is the speed of light \citep[see][]{bs10}. As shown in \citet{dh04}, we can write the Cartesian components of $\vec{B_{\rm ret}}$ as follows

\begin{eqnarray}
\label{bfield}
B_{\rm ret,x} = \frac{\mu}{r^5} ( 3xz\cos\alpha_{\rm A} + \sin\alpha_{\rm A} ( [(3x^2 - r^2) + 3xyr_{\rm n} + (r^2-x^2)r_{\rm n}^2] \cos(\Omega t - r_{\rm n}) \nonumber \\
 + [3xy-(3x^2-r^2)r_{\rm n} - xyr_{\rm n}^2] \sin(\Omega t - r_{\rm n}) ) ) \nonumber \\
B_{\rm ret,y} = \frac{\mu}{r^5} ( 3yz\cos\alpha_{\rm A} + \sin\alpha_{\rm A}([3xy + (3y^2-r^2)r_{\rm n} -xyr_{\rm n}^2 ] \cos(\Omega t - r_{\rm n}) \nonumber \\
+ [(3y^2-r^2) - 3xyr_{\rm n} + (r^2-y^2)r_{\rm n}^2 ] \sin(\Omega t - r_{\rm n}))) \\
B_{\rm ret,z} = \frac{\mu}{r^5} ( (3z^2-r^2)\cos\alpha_{\rm A} + \sin\alpha_{\rm A}[(3xz+3yzr_{\rm n} - xzr_{\rm n}^2) \cos(\Omega t - r_{\rm n}) \nonumber \\
+ (3yz - 3xzr_{\rm n} - yzr_{\rm n}^2) \sin(\Omega t - r_{\rm n})])~. \nonumber
\end{eqnarray}

\noindent
Here $r_{\rm n} \equiv r/R_{\rm LC}$, where $R_{\rm LC}$ is the light cylinder radius. Using pulsar A's spin period ($P_{\rm s}=22.7$ ms), we fix pulsar A's light cylinder radius to be $R_{\rm LC} = cP_{\rm s}/2\pi = 1100$~km in the calculation. Then the ratio $r_{\rm n}$ is small in the vicinity of the NS surface and the retarded field configuration is almost the same as the static field configuration in the `near' zone (i.e., $r << R_{\rm LC}$).
As explained in \citet{dh04}, the location of any corotating point within the magnetosphere does not depend on time. In other words, the retarded magnetic dipole field configuration is fixed in space and time in corotating frame. 
However, the field line structure rotates around the rotation axis as a whole with the pulsar spin.

As shown in \citet{fsk+13}, and also in Section~\ref{circular}, pulsar A's spin colatitude $\delta_{\rm A}$ is almost zero. Thus, to simplify the model, we assume that pulsar A's spin axis is aligned with the orbital angular momentum ($\delta_{\rm A}=0\degr$). In order to determine the magnetic field lines, we use Equation~(\ref{bfield}) with the fourth-order Runge-Kutta integration method. Two angles ($\theta_{\rm m}$, $\phi_{\rm m}$) are used to define the footpoint of the magnetic field line on the NS surface, where we assume a NS radius of R$_{\rm NS}=$10~km. Here, $\theta_{\rm m}$ is the colatitude angle from the magnetic axis and $\phi_{\rm m}$ is the azimuth of the field line footpoint. Then we determine the field line which starts from this initial point. 
First, we determine the last closed field lines, by varying $\theta_{\rm m}$ for a given $\phi_{\rm m}$ (i.e., bisection in $\theta_{\rm m}$ at fixed $\phi_{\rm m}$) until the field line becomes tangent to the light cylinder. 
We then define the PC region by calculating the footpoint of these last closed field lines on the NS surface. The shape of the PC region predicted by a retarded magnetic field is typically not symmetric around the magnetic axis and is dependent on $\alpha_{\rm A}$ \citep[see Figure 2 in][]{dh04}.     
As pointed out in \citet{dhr04}, we use bisection in $\phi_{\rm m}$ at fixed $\theta_{\rm m}$ around the `notch' region to correct the PC rim. 
A field line with a smaller $\theta_{\rm m}$ is open with respect to the light cylinder and referred as an open field line. We then model these open field lines which are located within the PC region. In order to do that, we define a set of field line footpoint rings within the PC region with a fixed colatitude ratio of $\theta_{\rm m} / \theta_{\rm rim}$, where $\theta_{\rm rim}$ is the colatitude of the PC rim of a given $\phi_{\rm m}$. We calculate footpoints with a fixed $4\degr$ increment in the azimuthal direction and obtain 90 field lines in each ring. Then, we define several sets of footpoint rings according to the colatitude ratio from $0.1$ to $1$ with an increment of $0.05$. 
We note that increasing the number of footpoints in a ring and the number of rings within the PC would smooth the modeled pulse profile and the shape variation becomes negligible. By testing several values, we found that the above given increments on these two parameters were sufficient for this analysis. Starting from these footpoints, we draw the open field lines using numerical integration.

Although the exact radio emission mechanism is not well understood, we believe that the charged particles stream along magnetic field lines and emit radiation tangential to the local magnetic field line at the emitting point. Therefore, we first determine the photon emission direction at any given emission point on a field line in the PC region. In order to do that, we perform numerical integration with a fixed step size. By using a smaller step size, we can safely assume that the unit vector of the field line segment at a given point is indeed tangent to the field line. This guarantees that the unit vector of emitted photons ($\hat{\eta}'$) are also tangential to the field line at this point. The unit vector of photons are represented by two angles: the colatitude of the tangent from the rotation axis or the viewing angle $\zeta$ and the azimuth angle or the spin longitude $\phi$. Here, we consider the inertial observer frame, where the direction of the photon is not $\hat{\eta}'$. In order to get the photon direction correctly in this frame ($\hat{\eta}$), we use the aberration formula \citep[see Equation (1) in][]{dr03} that accounts for the local corotational velocity with respect to the inertial observer frame as follows

\begin{equation}  
\hat{\eta} = \frac{\hat{\eta}' + [\gamma+(\gamma-1)(\vec{\beta}\cdot\hat{\eta}')/\beta^2]\vec{\beta}}{\gamma(1+\vec{\beta}\cdot\hat{\eta}')}~,
\end{equation}

\noindent
where $\gamma = (1-\beta^2)^{-1/2}$ is the Lorentz factor and $\vec{\beta} = (\vec{\Omega} \times \vec{r})/c$ is the local corotation velocity in units of the speed of light at the emission point $\vec{r}$.
Due to aberration, we observe emission slightly earlier in time, or in spin phase. 
The aberration is in particular important when the emission point $\vec{r}$ is large, i.e., the maximum aberration occurs close to the light cylinder radius with the maximum corotation velocity. Since pulsar A is nearly orthogonal, the corotation velocity of charged particles is important and hence, we include aberration in the model.
This is the aberration correction method given in many previous studies. However, \citet{bs10} pointed out that this method leads to an inconsistency since the retarded dipole field was traced in the lab frame, but the aberration was computed treating the field in the instantaneous corotating frame. They showed that this can be corrected by taking a coordinate transformation first, and then correct for the aberration. However, this is a second-order correction in $r/R_{\rm LC}$, so that it is not important for low-altitude radio emission and does not affect our results. Therefore, we do not include this second order correction in our model.

The next step is to include the photon propagation time delay between low- and high-altitude emission reaching the observer.
This delay is given as $\vec{r}\cdot\hat{\eta}/R_{\rm LC}$ and is added to the aberration corrected azimuth $\phi$ of the emission point to get the correct phase of each photon \citep[see][]{dr03}. One of the important observable consequences of the delay is that trailing photons are piled up at a particular spin phase, giving rise to large number of photons (`caustic' regions) and producing emission peaks at the line-of-sight as the pulsar rotates. Both aberration and propagation time delay are most important in outer magnetospheric models that describe high-altitude emission, but they cannot be neglected at low altitudes.

We then map aberration and propagation delay corrected photons in a parameter space of $\zeta$ versus $\phi$, which is usually called a sky map. 
We use a bin size of $1\degr$ in both $\zeta$ and $\phi$ directions.
We assume the coherent radio emission is generated at a particular height above the PC region (see Figure~\ref{diagram}). We further assume that the emissivity of the photon emission is constant across this region. The modeled pulse profile is generated by limiting the photon emission to a particular region at this height above the PC (see Section~\ref{fitting} for more detail). Then, a horizontal cut of the sky map at a given viewing angle $\zeta$ returns the model pulse profile. By fitting the model pulse profile to pulsar A's observed profile, we can determine the radio emission altitude $h$ and the size of the radio beam $\rho_{A}$ based on the last closed field lines. This $\rho_{A}$ is an independent estimate for pulsar A's beam size from what we determined in Section~\ref{circular} through a simple circular beam model.

\subsection{Fitting Pulse Profiles of Pulsar A}
\label{fitting}

\begin{figure*}
\epsscale{.50}
\plotone{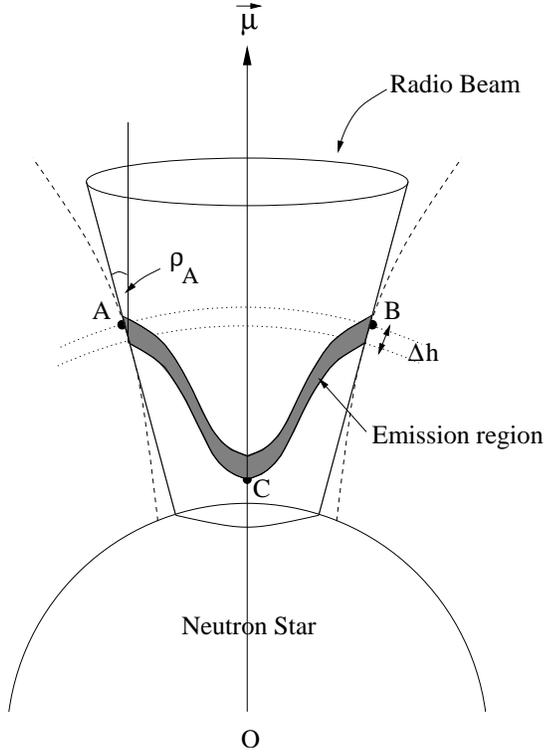}
\caption{
Schematic diagram of the radio emission beam (not in scale). Dashed lines are the two last closed magnetic field lines. The radio beam is tangent to these two field lines at point A and B. In other words, the radio beam is bounded by these two points. The coherent radio emission is generated within the filled area with thickness $\Delta h$. The colatitude of any given radio photon is less than the half-opening angle $\rho_A$ of the beam, i.e. all the photons are emitted within this boundary of the beam. The emission height at the edge of the beam ($h$) is shown as the length $OA$ ($= OB$). The emission height along the magnetic axis ($r_0$) is shown as the length $OC$. We assume the height of the emission region decreases exponentially from the beam edge towards the center.
\label{diagram}}
\end{figure*}

As the PC region is bounded by the last closed field lines, we assume the outer edges (i.e., 5\% intensity levels), or wings, of a pulse profile are generated from the emission within a thickness of $\Delta h$ along these last closed magnetic field lines at a emission altitude $h$ (see Figure~\ref{diagram}). The inner part of the pulse component is assumed to be  generated from the emission within the same thickness of $\Delta h$ along open field lines above the NS surface. 
If we fit the entire pulse profile including outer and the inner parts of the pulse component at once, the region around the peak of the profile dominates the result, providing unrealistically large beam sizes and emission altitudes. Thus, we fit pulse profiles in two steps to avoid this issue.

The first step (Step One) is to estimate the emission altitude $h$ and the emission width $\Delta h$ which correspond to the profile wings at 5\% intensity level. In order to do this, we map the photon emission from the last closed field lines by varying $h$ and $\Delta h$. Then we fit the modeled profile wings to the observed profile wings and obtain the best-fit $h$ and $\Delta h$ by a maximum likelihood method that we used in Section~\ref{circular}.
We determine the half-opening angle or the beam size $\rho_{\rm A}$ of the radio beam from the direction of the photon emission at this best-fit $h$.

The second step (Step Two) is to model the entire region of the open magnetic field lines fixing the emission altitude to be the best-fit $h$ and $\Delta h$ obtained from Step One. We then compare the entire model pulse profile with the observed profile. However, using a single emission height from the edge of the beam is unrealistic as the emission is not necessarily generated at one particular altitude across the entire open field line region \citep{lm88}. 
Therefore, we investigate different emission altitudes across the beam in addition to a constant emission altitude. In this model, we simply assume the radio emission altitude falls off exponentially with height towards the center of the beam from the edge (see Figure~\ref{diagram}).
Then we write an expression for the emission altitude at any point across the beam as $r=h\exp(-(\rho_{\rm r} - \rho_{\rm A})^2 /2\sigma^2)$. Again we assume that emission is generated within the thickness of $\Delta h$ at altitude $r$. We emphasize that $h$ is the emission height at the edge of the beam where the pulse profile wing is formed and is obtained from Step One. The parameter $\rho_{\rm r}$ is the colatitude of the photon at $r$ with respect to the pulsar's magnetic axis and can be obtained from the direction of the photon emission. This definition implies an inequality relation: $\rho_{\rm r} \leq \rho_{\rm A}$. The parameter $\sigma$ determines the shape of the cross section of the emission thickness and can be written as $\sigma = \rho_{\rm A} / \sqrt{-2\ln (r_0/h)}$, assuming $\rho_{\rm r} = 0\degr$ along the magnetic axis at lower altitude. The height $r_0$ is the emission altitude at the magnetic axis. Once $h$ and $\rho_{\rm A}$ are obtained from Step One, we vary $r_0$ and fit the full model pulse profile to the observed profile as explained in Step Two and estimate the best-fit $r_0$. Instead of assuming two identical beams, we assume the emission altitudes at wings of north ($h_{\rm N}$) and south ($h_{\rm S}$) beams can be different and calculate each separately. Likewise, we define the emission altitude at the magnetic axis from the two beams as $r_{\rm 0,N}$ and $r_{\rm 0,S}$.

Below is a summary of our prescription for the pulse profile fitting using the PC model: (1) model the field line structure based on a retarded vacuum dipole for a given magnetic inclination $\alpha_{\rm A}$ (see Equation~(\ref{bfield})); (2) model the photon emission from the open and last closed field lines including the effects of aberration and light propagation delay; (3) assume the radio emission is generated from these field lines within a thickness of $\Delta h$ at a given altitude above the PC region, map the photon emission from this region in the space of $\zeta$ vs $\phi$; (4) obtain a model pulse profile for the viewing angle of the line-of-sight $\zeta_{\rm E}$; (5) fit the model pulse profile to the observed one and constrain the radio beam geometry.

In this work, we obtain four sets of ${ h, \rho_{\rm A}, {\rm and}~ r_{\rm 0}}$ using results from the model CBM and constraints from \citet{gkj+13} combined with the PC model. The results are described in the next subsection, where subscripts of $N$ and $S$ denote the north and south poles.

\subsection{Results}

\begin{figure*}
\epsscale{.70}
\plotone{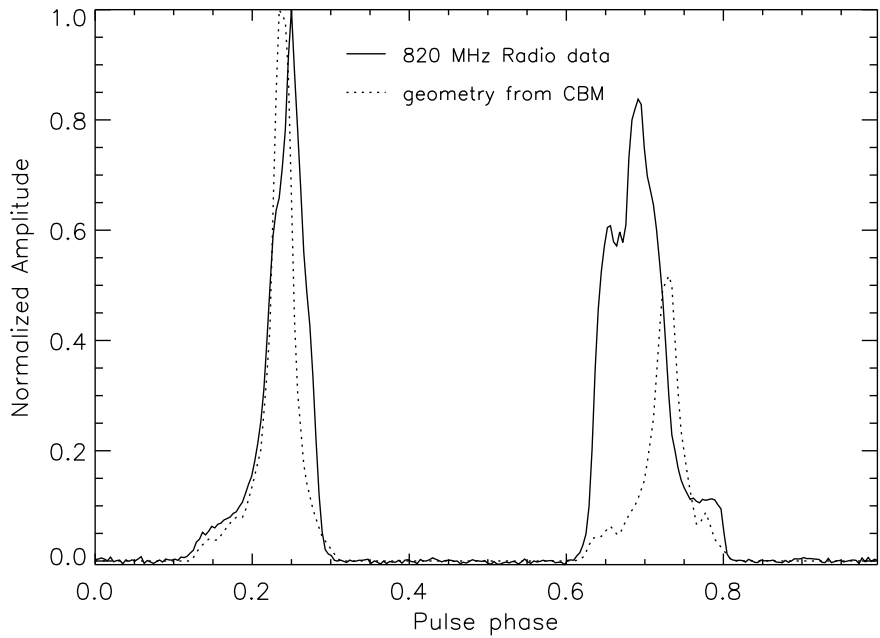}
\caption{
The modeled pulse profile based on the circular beam model ({\it dotted}). The observed integrated pulse profile at 820~MHz is shown as a {\it solid} curve. Note that the observed profile has 256 bins across the spin phase. All profiles are normalized to the brightest peak obtained by each model.
The emission altitude is estimated to be $\sim$10~R$_{\rm NS}$. 
\label{profile_part1}}
\end{figure*}

The geometry of CBM yields that $\zeta_{\rm E}$ is consistent with being constant in time,  due to an aligned or nearly aligned spin axis ($\delta_{\rm A} \leq 2\fdg8$). Therefore, a choice of MJD does not affect the model pulse profile significantly, if at all. As shown in previous studies and again confirmed in Section~\ref{data}, the pulse profiles of pulsar A do not show a significant time evolution. Therefore, we consider the observed pulse profile on MJD 53861 (Figure~\ref{profile}) as the time-independent observed pulse profile of A and obtained $h$ and $\Delta h$ by fitting the model pulse profile to this one.
According to the geometry, the best-fit radio beam parameters are estimated to be $h_{\rm N} = 10^{+1}_{-2}$~R$_{\rm NS}$, $h_{\rm S} = 11\pm2$~R$_{\rm NS}$, $\Delta h = 1\pm1$~R$_{\rm NS}$, 
$r_{0\rm,N} = 2^{+7}_{-1}$~R$_{\rm NS}$, and $r_{\rm 0,S} = 5^{+6}_{-4}$~R$_{\rm NS}$.
The beam half-opening angles are $\rho_{\rm N} = 31\degr \pm 1\degr$ and $\rho_{\rm S} = 33\degr \pm 1\degr$. 
Figure~\ref{profile_part1} presents the best-fit pulse profile obtained from the PC model with the geometry determined by CBM, overlaid with  A's pulse profile (solid).

\begin{figure*}
\epsscale{.70}
\plotone{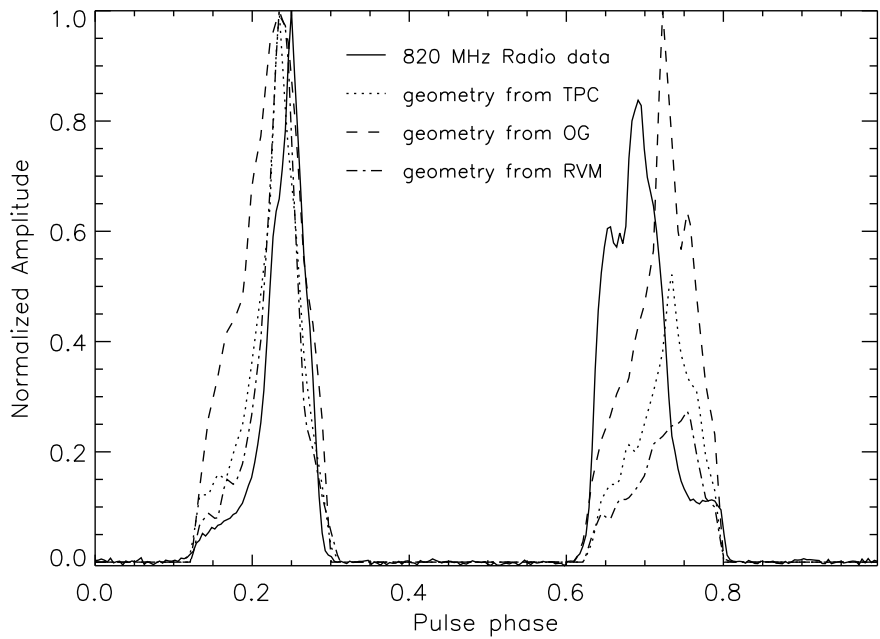}
\caption{
As described in Figure~\ref{profile_part1}, but for the geometry derived from TPC ({\it dotted}), OG ({\it dashed}), and radio polarization along with RVM ({\it dotted dashed}). The estimated emission altitudes from these three models range between 10--15~R$_{\rm NS}$.
\label{profile_part2}}
\end{figure*}

Due to the lack of evolution in A's pulse profile, \citet{gkj+13} set $\delta_{\rm A} = 0\degr$ and 
fit TPC and OG emission models to gamma-ray light curves, separately, in order to obtain the geometry of the pulsar (see Table~1).
The best-fit results from the TPC model are $\alpha_{\rm A} = 80(9)\degr$ and $\zeta_{\rm E} = 86(14)\degr$. With these parameters, we apply our PC beam model to search for the radio emission altitude and the radio beam size. The emission heights are estimated to be $h_{\rm N} = 11^{+1}_{-2}$~R$_{\rm NS}$ and $h_{\rm S} = 11^{+2}_{-1}$~R$_{\rm NS}$ with $\Delta h = 1.2^{+0.6}_{-0.9}$~R$_{\rm NS}$, $r_{0\rm ,N} = 2^{+8}_{-1}$~R$_{\rm NS}$, and $r_{\rm 0,S} = 1$~R$_{\rm NS}$ (i.e., on the NS surface) with an upper bound error of 1~R$_{\rm NS}$. The half-opening angles are $\rho_{\rm N} = 32\degr \pm 1\degr$ and $\rho_{\rm S} = 32\degr \pm 1\degr$. 
The OG model (i.e., $\alpha_{\rm A}=88(17)\degr$ and $\zeta_{\rm E}=74(14)\degr$) gives the best-fit parameters as follows: $h_{\rm N} = 12^{+2}_{-1}$~R$_{\rm NS}$ and $h_{\rm S} = 15\pm2$~R$_{\rm NS}$ with $\Delta h = 2\pm2$~R$_{\rm NS}$, $r_{\rm 0,N} = 2^{+9}_{-1}$~R$_{\rm NS}$, and $r_{\rm 0,S} = 1$~R$_{\rm NS}$ (i.e., on the NS surface) with an upper bound error of 12~R$_{\rm NS}$. The half-opening angles with the OG model are estimated to be $\rho_{\rm N} = 33\degr \pm 2\degr$ and $\rho_{\rm S} = 38\degr \pm 1\degr$.
Figure~\ref{profile_part2} shows the comparison between  A's pulse profile (solid) with the best-fit pulse profiles obtained from  the PC model with the geometry estimated by the TPC and OG magnetosphere models, respectively.
By comparing our PC beam model pulse profile with the hollow-cone beam model profile given in \citet{gkj+13} (see Figure~2 therein), it is clearly seen that the PC beam model fits the pulse profile wings of both pulse components better. The brightest pulse of the modeled pulse profiles from both models are consistent with the observed profile. However, due to the hollow-cone nature of the radio beam model given in \citet{gkj+13}, the shape of P2 from the TPC-model-resulted pulse profile has a broad double peak structure. In contrast, our PC model results in a single peak structure for P2 and is more similar to the observed pulse component than the double peaked structure.

Lastly, if the constraints from polarization measurements provided by \citet{gkj+13}, $\alpha_{\rm A} = 99(8)\degr$ and $\zeta_{\rm E}=96(13)\degr$, are included with the PC model, we obtain the emission heights to be $h_{\rm N} = 10^{+1}_{-2}$~R$_{\rm NS}$ and $h_{\rm S} = 12^{+3}_{-1}$~R$_{\rm NS}$ with $\Delta h = 1.25\pm 1.0$~R$_{\rm NS}$, $r_{\rm 0,N} = 2\pm 1$~R$_{\rm NS}$, and $r_{\rm 0,S} = 1$~R$_{\rm NS}$ with an upper bound error of 1~R$_{\rm NS}$. The half-opening angles of the two beams are estimated to be $\rho_{\rm N} = 30\degr \pm 1\degr$ and $\rho_{\rm S} = 33\degr \pm 1\degr$, respectively. 
Figure~\ref{profile_part2} shows the best-fit pulse profile obtained from  the PC model with the geometry estimated by the polarization observation.
As shown in \citet{gkj+13}, the polarimetric profile for pulsar A is complicated. They fit their model only to the rapid polarization angle sweep seen in the linearly polarized region around the peak of each pulse component and then constrained the geometry, and the emission altitudes based on the phase lag between the polarization angle sweep and the magnetic axis. 
Their results showed that the emission altitude corresponding to the central region of the second brightest pulse component (i.e., P2 in this work) is $\sim$1~R$_{\rm NS}$, which is in agreement with our result (i.e., $r_{\rm 0,S} = 1$~R$_{\rm NS}$). However, their best-fit emission altitude corresponding to the central region of the brightest pulse component is $\sim$12~R$_{\rm NS}$, which is greater than our best-fit value of $r_{\rm 0,N} = 2$~R$_{\rm NS}$. 
We note that the geometry and the emission altitudes constrained by \citet{gkj+13} were meaningful only for small sections of the pulse profile, close to the peak of each pulse component.  
However, we fit the PC model for the entire pulse profile assuming this geometry. This inconsistency could be the reason for the differences in resulting emission altitudes for P1 from the two models.


\section{The Orbital Geometry of the Double Pulsar}
\label{full_geo}

When the beam geometry of A and B pulsars are known, it is possible to calculate the relative angle ($\Delta_{\rm S}$) between the spin axes of the two pulsars and to fully configure the orbital geometry of the Double Pulsar.
The main results for A's geometry are summarized in Table 1. 
\citet{pmk+10, plg+12} presented the geometry of pulsar B and \citet{kpm13} provided constraints on the beaming fraction.
 
Utilizing the results described in earlier sections, we can write $\Delta_{\rm S}(t)$ as follows

\begin{equation}
\label{re_angle}
\cos{(\Delta_{\rm S}(t))} = \cos\delta_{\rm A} \cos\delta_{\rm B} + \sin\delta_{\rm A} \sin\delta_{\rm B} \cos(\Delta \phi_{\rm prec}(t)),
\end{equation}

\noindent
where $\Delta_{\rm S}(t)$ is the relative angle between the spin axes of A and B at time $t$. The angles $\delta_{\rm A}$ and $\delta_{\rm B}$ are spin misalignment angles of A and B with respect to the orbital angular momentum. 
The angle $\Delta \phi_{\rm prec}(t)$ is the relative spin precession angle and is defined by $\Delta \phi_{\rm prec}(t) = \phi_{\rm prec,A}(t) - \phi_{\rm prec,B}(t)$, where $\phi_{\rm prec,i}(t) = \Omega_{\rm prec,i}(t-T_0)$ is the spin precession phase and $\Omega_{\rm prec,i}$ is the spin precession rate for ${\rm i}=$ A and B pulsars. 
Note that the angle $\Delta_{\rm S}(t)$ is not affected by the details of our assumptions on the pulsar radio beams or magnetic misalignment. 
Geodetic precession of the two pulsars would cause $\Delta_{\rm S}$ to change over time.

With the particular geometric framework, the minimum and maximum $\Delta_{\rm S}(t)$ are given by $\delta_{\rm B}-\delta_{\rm A}$ and $\delta_{\rm B}+\delta_{\rm A}$, respectively. Based on our results and \citet{plg+12}, $\delta_{\rm B} > \delta_{\rm A}$ (Table 1). 
At the current epoch, $\Delta_{\rm S}(t)$ for CBM is $138(5)\degr$.
Considering the 2$\sigma$ uncertainties of $\delta_{\rm A}$ and $\delta_{\rm B}$ given in Table~1, we estimate the uncertainty of $\Delta_{\rm S}(
t)$ to be $\pm$$6\degr$.
If $\delta_{\rm A}$ does not equal 0, then $\Delta_{\rm S}$ will show a variation with time with a period of 1385 years, based on the precession periods of pulsar A (75 yrs) and pulsar B (71 yrs). 
The evolution of $\Delta_{\rm S}$ of the Double Pulsar is expected to follow Equation~(\ref{re_angle}) until tidal interactions become important in $\sim$85~Myr at merger phase \citep[see][]{kpm13}.

\section{Discussion and Conclusions}
\label{dis}

In this work, we constrain A's beam geometry assuming a double-pole geometry to
 estimate the magnetic misalignment angle ($\alpha_{\rm A}$) and the colatitude of the spin axis ($\delta_{\rm A}$).
\citet{fsk+13} estimated these angles to be $\alpha_{\rm A} = 90\degr \pm 16\degr$ and $\delta_{\rm A} \le 2\fdg3$ at 68\% confidence. Our results have smaller errors, but  are consistent with these (see Table~1).
As shown in \citet{mkp+05}, \citet{fsk+08,fsk+13} and Section~\ref{data} in this work, the orthogonal geometry of the pulsar provides no secular variation in the observed pulse profiles. 
In the studies of other pulsar binaries given in Table~1, similar geometrical frameworks have been used to constrain the geometry based on the observed pulse profile variations due to spin precession. The pulse profile evolution of these systems are all detectable due to their large spin misalignment. However, the evolution is somewhat long-term due to the much smaller spin precession rates compared to that of the double pulsar; see \citet{kra98} for the details of the long-term profile evolution of PSR B$1913+16$. As shown for pulsar A, we note that the pulsar geometry, and the spin precession rate, both play an important role in the profile evolution of binary pulsars.

The recent {\it Fermi} detection of pulsed gamma-ray emission from pulsar A revealed that the peaks of the high-energy and radio profiles are not aligned in spin phase \citep{gkj+13}. This implies that high- and low-energy emission is produced at two different locations in the magnetosphere.
\citet{gkj+13} used the OG and TPC models  to describe the high-energy emission and gamma-ray light curves from pulsar A. 
The RVM, based on the radio polarization measurement, is used to constrain A's beam geometry.
We incorporate $\alpha_{\rm A}$ and $\zeta_{\rm E}$ from these models and the model given in Section~\ref{circular} in our PC  magnetosphere beaming model to synthesize the pulse profile of A and then estimate the radio beam size and the emission altitude. 
All our model pulse profiles are qualitatively in agreement with A's observed profile, but none of them are perfect fits. We find that including the aberration effects results in a better fit to the leading step-like part of the P1 component, but the peak of P2 component does not fit well with observation. 
However, the edges of P2 fit reasonably well with the observed pulse component.
Thus, a high photon density towards the leading edge of the south emission beam compared to its trailing edge can result in more photons around the leading edge of P2 component. This may move the peak of P2 to lower spin phases and fit with the observed peak better. In reality, different beam shapes can exist; two asymmetric beams may provide a better fit for observation. Further, a partially filled patchy beam structure can be another option in pulsar beam modeling. 
However, these beam structures are not easily modeled due to the large number of free parameters associated with them.

Using pulsar A's spin period, period derivative, and the observing frequency of 820~MHz in the expression derived by \citet{kg03a}, we obtain A's emission altitude ($h_{\rm A}\sim$$10$~R$_{\rm NS}$). Empirical fits to canonical, non-recycled, pulsars imply a correlation between the radio beam size and the pulsar's spin period: $\rho = 5\fdg4 P_{\rm s}^{-0.5}$ \citep{kxl+98}. It is not clear whether the recycled pulsars follow this empirical relation, but if we assume this, the beam size of A is estimated to be $\sim38\degr$. In this case, $h_{\rm A}$ and $\rho_{\rm A}$ obtained from the PC beam model for pulsar A are consistent with those predicted by the above relations. 
Regardless of assumption on the $\rho-P_{\rm s}$ relation, the emission altitudes for recycled pulsars J0437--4715 ($P_{\rm s}=5.75$~ms) and B1913$+$16 ($P_{\rm s}=59$~ms) were estimated to be $h<$9~R$_{\rm NS}$ \citep{gt06} and $h<$20~R$_{\rm NS}$ \citep{kg97}, respectively.
Our results on pulsar A's emission altitudes ($<$15~R$_{\rm NS}$) are consistent with those of above two pulsars and the differences in estimates are associated with different magnetosphere sizes ($\equiv cP_{\rm s}/2\pi$), based on their pulse periods.

Our measurement of $\Delta_{\rm S}=138^{\circ}$, at the current moment, is the first measurement of the spin orientation in a NS-NS binary. This information is useful in studying the final evolution of NS-NS binaries. 
Since the detection of gravitational waves from merging binary neutron stars is a major goal of gravitational-wave observatories such as Advanced LIGO \citep{har10} and Advanced Virgo \citep{daa+13}, constraining this parameter for known or new NS-NS binaries in the future is important.
Recently, \citet{brown12} showed that including spin effects in GW search templates could provide increases in sensitivity. They also discussed that GWs from all known NS-NS binaries in the Galactic disk can be described with non-spinning waveforms.
Even for the Double Pulsar, with the fastest spinning recycled pulsar among the known NS-NS binaries, the aligned spin with respect to the orbital normal and the long spin period of B make a non-spinning waveform a suitable template to search for GWs. However, spin effects are likely to be more important for NS-NS binaries with two faster-spinning neutron stars, so that our measurement provides a useful path to study such complicated systems.

\bigskip

\acknowledgments
BBPP, MAM, and CK are supported through the Research Corporation. CK is supported in part by the National Research Foundation Grant funded by the Korean Government (No. NRF-2011-220-C00029). CK thank Hee-Il Kim and Alex Nielsen for useful discussions. Pulsar research at UBC is supported by an NSERC Discovery grant.

\bibliography{psrrefs,modrefs,journals,0737Ack} 
\bibliographystyle{mn2e}

\end{document}